\documentclass[a4paper,11pt]{article}
\pdfoutput=1 

\usepackage{jheppub1} 

\usepackage[T1]{fontenc} 
\usepackage{float}
\usepackage{epstopdf}
\usepackage{amsmath}

\title{\boldmath Magnetized topological black holes of dimensionally continued gravity}

\author{Askar Ali and Khalid Saifullah} 
\affiliation{Department of Mathematics, Quaid-i-Azam University, Islamabad, Pakistan}

\emailAdd{askarali@math.qau.edu.pk} \emailAdd{ksaifullah@fas.harvard.edu}

\abstract{In this paper, a large family of topological black hole solutions of dimensionally continued gravity are derived. The action of Lovelock gravity is coupled to the exponential electrodynamics and the equations of motion are solved in the presence of a pure magnetic source.  We work out the metric functions in terms of the parameter $\beta$ of exponential electrodynamics, and magnetic charge. Further, we couple Lovelock gravity to power-Yang-Mills theory and construct black holes, in diverse dimensions, having Yang-Mills magnetic charge. We also discuss the asymptotic bahaviour of metric functions and curvature invariants at the origin for both the models. The thermodynamics of resulting magnetized black hole solutions in the framework of two different models is also studied. The thermodynamical quantities like Hawking temperature, entropy and specific heat capacity at constant charge are found and we show that the resulting quantities satisfy the first law of black hole thermodynamics. We also study the magnetized hairy black holes of dimensionally continued gravity. 
\vspace{55 mm}
}

\notoc 

\begin{document}
\maketitle
\flushbottom 


\section{Introduction}
\label{sec:intro}
Since the theory of general relativity is nonrenormalizable, therefore, higher derivative gravitational theories have been investigated, because the higher derivative corrections to the familiar Einstein's theory produce a power-counting renormalizable theory \cite{1}. Further, modifying gravity gives an alternative way to understand the acceleration and expansion of the universe without the introduction of dark energy in the model. Thus, this is another reason to study higher derivative theories. Among the different higher derivative modified theories, Lovelock gravity \cite{2}, which contains dimensionally continued Euler characteristics, has a unique property that, in four dimensions, it reduces to general relativity. The field equations corresponding to the Lovelock gravity contain only the metric and its first two derivatives, thus the linearized form of this theory is free of ghosts. The second order Lovelock gravity, i.e., the Gauss-Bonnet gravity emerges in string theory in the low energy limit \cite{3,4}. However, due to the presence of a lot of Lovelock coefficients in Lovelock gravity, it is very difficult to interpret the physical meaning of the solution, therefore Banados, Teitelboim and Zanelli \cite{5} proposed a suitable choice of these coefficients which allows us to write the solution in an explicit form. This theory of gravity that is obtained from this particular choice of Lovelock coefficients is known as dimensionally continued gravity (DCG). In the literature \cite{5, 6, 7} neutral and charged black hole solutions of DCG have been studied. The hairy black holes of DCG have also been constructed recently \cite{8,9}.

It has been shown that the standard Maxwell's theory is not always workable for studying electromagnetic fields. In 1934 nonlinear electrodynamics was proposed by Born and Infeld which has the property of cancelling the divergences of electron's self-energy \cite{10}. Later in 1936, Heisenberg and Euler also put forward a nonlinear electromagnetic theory to explain the quantum electrodynamics phenomena \cite{11}. The Born-Infeld electrodynamics could also be reproduced in the framework of string theory \cite{12}. The action of Born-Infeld theory also governs the dynamics of D3-branes \cite{13}. Other theories, that have recently been found from a particular form of Born-Infeld theory, include Dirac-Born-Infeld inflation theory and Eddington-inspired Born-Infeld theory \cite{14,15}. These theories have also been used in the study of dark energy, holographic entanglement entropy and holographic superconductors \cite{16,17,18}. The first black hole solution of Einstein's theory with Born-Infeld electrodynamical source was given by Hoffmann \cite{185}. Subsequently, many spherical black hole solutions which are asymptotic to the Reissner-Nordstr\"{o}m solution were derived with other nonlinear electrodynamical sources \cite{19,20}. The exponential electrodynamics model \cite{21} is also used to construct the asymptotic Reissner-Nordstr\"{o}m black hole solution having magnetic charge.

Black hole solutions of modified gravities have also been studied in the framework of nonlinear electromagnetic theory. For example, nonsingular magnetically charged black hole solutions \cite{AS} and electrically charged black hole solutions of DCG have been constructed \cite{22} where exponential and Born-Infeld electromagnetic fields are taken as sources. In the present work we derive magnetized black hole solutions of DCG coupled with exponential electrodynamics and then generalize our solution to black holes which contain scalar hair.

More recently black hole solution has been found where the nonlinear electromagnetic source is expressed in powers of Maxwell's invariant $(F_{\mu\nu}F^{\mu\nu})^q$, where $q$ is an arbitrary real number \cite{23}. The nonlinearity involved in this power-Maxwell theory is radically different from the familiar Born-Infeld electrodynamics. One property of power-Maxwell formalism is that for the special case of $q=d/4$, where $d$ represents the dimension of spacetime, it yields the traceless matter tensor which indicates the satisfaction of conformal invariance. Instead of considering the power-Maxwell theory, we investigate black hole solutions with power-Yang-Mills source \cite{24} in this paper, that is, we can choose the source of gravity as $(F^{(a)}_{\mu\nu}F^{(a)\mu\nu})^q$, where $F^{(a)}_{\mu\nu}$ denotes the Yang-Mills field with $1\leq a\leq \frac{1}{2} (d-1)(d-2)$ and $q$ is a real number. In Ref. \cite{24} Lovelock black holes with a power-Yang-Mills source have been studied and magnetically charged solutions are obtained. In this paper we construct black hole solutions and hairy black hole solutions of DCG in the presence of  power-Yang-Mills field.

Thermodynamics of black holes is an interesting aspect of the subject that attracts much attention not only in linear Maxwell's theory but also in the nonlinear theories \cite{25,26}. For instance, thermodynamics of spherically symmetric black holes with exponential electromagnetic source has been discussed \cite{21}. Similarly, thermodynamics of Lovelock black holes in the framework of power-Yang-Mills theory has been studied \cite{24}. In our work we also study thermodynamics of the resulting black hole solutions in DCG within the two different models.

The plan of the paper is as follows. In Section 2, the exponential electrodynamics model is coupled with Lovelock gravity and we find a family of magnetized black hole solutions of DCG depending on the parameter $\beta$ of exponential electrodynamics. In this section the metric functions are calculated and we study thermodynamics of black holes with exponential electrodynamical source. We also work out the hairy black hole solution of DCG within this model. In Section 3, we couple Lovelock gravity to power-Yang-Mills theory and obtain the magnetized black hole solutions with and without scalar hair of DCG. We also study thermodynamics of these black hole solutions. Section 4 deals with the Yang-Mills hierarchies. Finally we conclude our results in Section 5.

\section{Topological black holes of DCG coupled to exponential electrodynamics} 

\subsection{Magnetized black hole solution}

The action function for Lovelock gravity with exponential electromagnetic source \cite{21, 22} in diverse dimensions is written in the form
\begin{equation}
I=\frac{1}{16\pi G}\int d^dx\sqrt{-g}\bigg[\sum_{p=0}^{n-1}\frac{\alpha_p}{2^p}\delta^{\mu_1...\mu_{2p}}_{\nu_1...\nu_{2p}} R^{\nu_1\nu_2}_{\mu_1\mu_2}...R^{\nu_{2p-1}\nu_{2p}}+L(\digamma)\bigg],
\label{1}
\end{equation}
where
\begin{equation}
L(\digamma)=-\digamma\exp{(-\beta\digamma)}, \label{2}
\end{equation}
 and  $\digamma=F_{\mu\nu}F^{\mu\nu}=(\textbf{B}^2-\textbf{E}^2)/2$ , $\beta$ is the parameter of our model, $\textbf{B}$ is the magnetic field and $\textbf{E}$ is the electric field. Here $G$ is the Newtonian constant, $\delta^{\mu_1...\mu_{2p}}_{\nu_1...\nu_{2p}}$ is the generalized Kronecker delta of order $2p$ and $d=n+2$ where $n$ is a natural number. The coefficients $\alpha_p$ in (\ref{1}) represent arbitrary constants, however, in the particular case of DCG, they are chosen in the form
 \begin{equation}
 \alpha_p=C^{n-1}_{p}\frac{(d-2p-1)!}{(d-2)!l^{2(n-p-1)}},\label{3}
 \end{equation}
where $l$ is related to cosmological constant. It is worth noting that DCG becomes Born-Infeld theory in even dimensions and Chern-Simons theory in odd dimensions \cite{5,6}. Varying (\ref{1}) with respect to the electromagnetic potential $A_{\mu}$ yields the equations of motion of nonlinear electromagnetic field
\begin{equation}
\partial_{\mu}\bigg[\sqrt{-g}F^{\mu\nu}\bigg(1-\beta\digamma\bigg)\exp{(-\beta\digamma)}\bigg]=0.
\label{4}
\end{equation}
After taking the variation of the action (\ref{1}) with respect to the metric tensor, $g_{\mu\nu}$, one can arrive at the equations of gravitional field
\begin{equation}
\sum_{p=0}^{n-1}\frac{\alpha_p}{2^{p+1}}\delta^{\nu\lambda_1...\lambda_{2p}}_{\mu\rho_1...\rho_{2p}} R^{\rho_1\rho_2}_{\lambda_1\lambda_2}...R^{\rho_{2p-1}\rho_{2p}}=T^{\nu}_{\mu},
\label{5}
\end{equation}
where $T^{\nu}_{\mu}$ represents the matter tensor given by
\begin{equation}
T^{\nu}_{\mu}=\exp{(-\beta\digamma)}\bigg[(\beta\digamma-1)F^{\mu\lambda}F_{\lambda}^{\nu}+\digamma g^{\mu\nu}\bigg].
\label{6}
\end{equation}
One can easily find the trace of the matter tensor in the form
\begin{equation}
T=-4\beta\digamma^2\exp{(-\beta\digamma)}, 
\label{7}
\end{equation}
which implies that for this model, the scale invariance is completely broken down, however, in the limit $\beta\rightarrow0$ one comes to the classical Maxwell's electrodynamics because the trace of the matter tensor becomes zero.

 The causality principle \cite{20,21} says that the group velocity of excitations in the background should not exceed the speed of light or we can say that tachyons will not appear. This principle will hold if $\partial L/\partial\digamma\leq 0$. Now from (\ref{2}) we have
  \begin{equation}
  \frac{\partial L}{\partial \digamma}=(\beta\digamma-1)\exp{(-\beta\digamma)}, \label{A}
  \end{equation}
which shows that causality is satisfied for our model if $(\beta\digamma-1)\leq0$. This gives us the requirement for the case of pure magnetic field
\begin{equation}
  r\leq \bigg(\frac{2}{\beta Q^2}\bigg)^{\frac{1}{2d-4}}. 
\label{B}
\end{equation}
According to the unitarity principle the norm of every elementary vacuum excitation will be positive definite, which means $(\partial L/\partial\digamma+2\digamma \partial^{2}L/\partial\digamma^2) \leq0$ and $\partial^{2}L/\partial\digamma^2\geq0$ \cite{21}. Thus, this principle holds if and only if
\begin{equation}
r\leq \bigg(\frac{0.44}{\beta Q^2}\bigg)^{\frac{1}{2d-4}}. 
\label{C}
\end{equation}
The weak energy condition (WEC) is fulfilled if and only if the relations $\rho\geq0$ and $(\rho+P_m)\geq0$, where $m=1,2,3...,d-1$ are satisfied. Note that, $\rho$ represents the energy density and $P_m$ are the principle pressures for each $m$. Thus, using matter tensor (\ref{6}) it can easily be verified that WEC is satisfied if and only if (\ref{B}) holds. From Eq. (\ref{6}) it is also easy to verify that the dominant energy condition (DEC) and strong energy condition (SEC) are also satisfied for our model if and only if 
 \begin{equation}
 r^{2d-4}\leq \mid\frac{2}{\beta Q^2}\mid. 
 \label{D}
 \end{equation}
 Note that, DEC guarantees that the speed of sound does not exceed the speed of light, while SEC says that there cannot be acceleration of the universe in the framework of exponential electrodynamics coupled with DCG.
 
 Since the action of Lovelock gravity (\ref{1}) is the sum of the dimensionally extended Euler densities, it is shown that there are no more than second order derivatives with respect to the metric tensor in its equations of motion. Moreover, the Lovelock gravity has been shown to be a ghostfree theory when expanding on a flat space, evading any problems with unitarity \cite{B1}. Here, note that although the Lagrangian density corresponding to Lovelock gravity contains some curvature terms with higher order derivatives, in essence it is not a higher derivative gravity theory because its equations of motion do not contain terms higher than second derivatives of the metric. From this it becomes clear that the Lovelock gravity is a ghostfree theory \cite{B2}. The action function (\ref{1}) corresponding to DCG looks very complicated due to the presence of so many terms. However, the static spherically symmetric black hole solutions can indeed be found \cite{B3} with the help of a real root of the corresponding polynomial equation. Since the gravity (\ref{1}) contains many arbitrary coefficients $\alpha_n$ it is not an easy task to extract physical information from the solution. Some authors \cite{5,6} choose a special set of coefficients which make the metric function simpler. These solutions could be explained as spherically symmetric solutions. Further, the black hole solutions having nontrivial horizon topology in this gravity with the special choice of coefficients have been studied \cite{7,B7}.

Here we determine the magnetically charged static black hole solution of DCG. For this we assume a pure magnetic field such that $\textbf{E}=0$, which yields from Maxwell's invariant, the form $\digamma=(B(r))^2/2=Q^2/2r^{2d-4}$, where $Q$ represents the magnetic charge. Now, for the general case of arbitrary coefficients $\alpha_n$, the static and spherically symmetric line element \cite{B8} was obtained in the form
\begin{equation}
ds^2=-f(r)dt^2+\frac{dr^2}{f(r)}+r^2 (h_{ij}dx^idx^j),
\label{8}
\end{equation}
where $h_{ij}dx^idx^j$ is the metric of $n$-dimensional hypersurface having constant curvature. Now using the above line element and substituting (\ref{6}) in Eq. (\ref{5}), the equation of motion becomes \cite{B3, B9}
\begin{equation}
\frac{d}{dr}\bigg[\sum_{p=0}^{n-1}\frac{\alpha_p(d-2)!}{2(d-2p-1)!}r^{d-1}\bigg(\frac{k-f(r)}{r^2}\bigg)^p\bigg]=32\pi G\exp{\bigg(\frac{-\beta Q^2}{2r^{2d-4}}\bigg)}\bigg(\frac{\beta Q^4}{2r^{4d-8}}-\frac{Q^2}{2r^{2d-4}}\bigg),
\label{9}
\end{equation}
where $k=0,1,-1$ associated to the codimensions-2 hypersurface with planar, spherical and hyperbolic topology respectively. If we chose $\alpha$ from Eq. (\ref{3}), that is, for the case of DCG, the above equation becomes \cite{B3, B9}
\begin{equation}
\frac{d}{dr}\bigg[r^{d-1}\bigg(\frac{1}{l^2}+\frac{k-f(r)}{r^2}\bigg)^{n-1}\bigg]=64\pi G\exp{\bigg(\frac{-\beta Q^2}{2r^{2d-4}}\bigg)}\bigg(\frac{\beta Q^4}{2r^{4d-8}}-\frac{Q^2}{2r^{2d-4}}\bigg).
\label{10}
\end{equation}
Integration of the above equation with respect to $r$ yields
\begin{align}
\label{11} 
f(r)&=\frac{r^2}{l^2}+k-r^2\bigg[\frac{16\pi G m}{r^{d-1}\Sigma_{d-2}}+\frac{\delta_d}{r^{d-1}}+\frac{\pi G Q^{\frac{1}{d-2}}\beta^{\frac{5-2d}{2d-4}}}{(d-2)2^{\frac{21-10d}{2d-4}}r^{d-1}}\bigg(\Gamma{\bigg(\frac{2d-5}{2d-4},\frac{\beta Q^2}{2r^{2d-4}}\bigg)} \nonumber \\ &-2\Gamma{\bigg(\frac{4d-9}{2d-4},\frac{\beta Q^2}{2r^{2d-4}}\bigg)}\bigg)\bigg]^{\frac{1}{d-3}},
\end{align}
where $m$ is a constant of integration which is associated to the ADM mass of black hole, $\Sigma_{d-2}$ represents the volume of $n$-dimensional hypersurface. The reason for the appearance of additional constant $\delta_d$ in Eq. (\ref{11}) is that one can expect the horizon of the black hole to shrink to a single point when $m\rightarrow0$ and the function $\Gamma{(s,x)}$ is the incomplete gamma function.

It is worth mentioning here that black strings and black branes can be thought of generalizations of black holes and they play a significant role in the AdS/CFT correspondence. It has been rightly pointed out \cite{CO} that while it is easy to construct black string and black brane solutions in the vacuum Einstein gravity by adding Ricci flat directions to Schwarzschild and Kerr solutions, it is nontrivial to obtain such solutions in Lovelock gravity, and numerical and other techniques have been used to construct such solutions. 

\begin{figure}[h]
	\centering
	\includegraphics[width=0.8\textwidth]{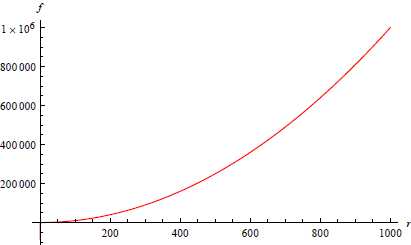}
	\caption{Plot of function $f(r)$ [Eq. (\ref{11})] vs $r$ for fixed values of $m=10$, $Q=0.01$, $\beta=0.01$ and $l=1$.}\label{Khan2}
\end{figure}

Now we discuss the asymptotic behaviour of the metric function at $r=0$. We take $k=1$ here (the cases $k=0,-1$ can be studied in a similar manner). Thus for even-dimensional spacetime, the asymptotic value of the metric function becomes
\begin{align} 
f(r) & =1+\frac{r^2}{l^2}-m^{\frac{1}{2s-3}}\bigg[r^{2k-5}+\frac{\delta_{d,2s}}{m(2s-3)}r^{2s-5}+\frac{\pi G}{m(2s-3)(2s-2)}\exp{\bigg(\frac{-\beta Q^2}{2r^{4(s-1)}}\bigg)} \nonumber  \\ & \times \bigg(\frac{32r^{4s-4}(3-2s)}{\beta(2s-2)}-\frac{Q^{\frac{1}{2s-2}}2^{\frac{20s-21}{4s-4}}\beta^{\frac{5-4s}{4s-4}}r^{6s-8}}{4s-4}-\frac{Q^2 2^{\frac{14s-15}{2s-2}}}{r^2s}+O(r^{10s-12})\bigg)\bigg], \nonumber  \\ & \hspace{70mm}  d=2s, s=1,2,3....
\label{12} 
\end{align}
Similarly, for odd-dimensional spacetime we have 
\begin{align}
f(r) & =1+\frac{r^2}{l^2}-m^{\frac{1}{2s-2}}\bigg[r^{2s-4}+\frac{\delta_{d,2s+1}}{2m(s-1)}r^{2s-4}+\frac{\pi G}{2m(s-1)(2s-1)}\exp{\bigg(\frac{-\beta Q^2}{2r^{4s-2}}\bigg)} \nonumber \\ & \times \bigg(\frac{64r^{2s-3}(1-s)}{\beta(2s-1)}-\frac{Q^{\frac{1}{2s-3}}2^{\frac{20s-11}{4s-2}}\beta^{\frac{3-4s}{4s-2}}r^{6s-5}}{4s-2}-\frac{Q^2 2^{\frac{14s-8}{2s-1}}}{r^{2s+1}}+O(r^{10s-7})\bigg)\bigg], \nonumber \\ & \hspace{70mm} d=2s+1, s=1,2,3.... \label{a} 
\end{align}
From the above asymptotic expressions, it is clearly seen that for the case of even-dimensions, metric function $f(r)$ is finite as $r\rightarrow 0$ for all values of $s\geq3$. In the case of odd-dimensions, $f(r)$ yields finite value for all $s\geq2$ at the origin, while at $s=1$ the function is infinite.
The event horizons can be found by solving the equation $f(r_{h})=0$. From Eq. (\ref{11}), we can write the ADM mass of the black hole in the form as
\begin{align}
m&=\frac{r_{h}^{d-1}\Sigma_{d-2}}{16\pi G}\bigg(\frac{1}{l^2}+\frac{k}{r_h^2}\bigg)^{d-3}-\frac{\delta_d}{16\pi G} \nonumber \\ &-\frac{Q^{\frac{1}{d-2}}\Sigma_{d-2}\beta^{\frac{5-2d}{2d-4}}}{(d-2)2^{\frac{5-2d}{2d-4}}} \bigg[\Gamma{\bigg(\frac{2d-5}{2d-4},\frac{\beta Q^2}{2r_h^{2d-4}}\bigg)}-2\Gamma{\bigg(\frac{4d-9}{2d-4},\frac{\beta Q^2}{2r_h^{2d-4}}\bigg)}\bigg]. \label{13}
\end{align}
\begin{figure}[h]
	\centering
	\includegraphics[width=0.8\textwidth]{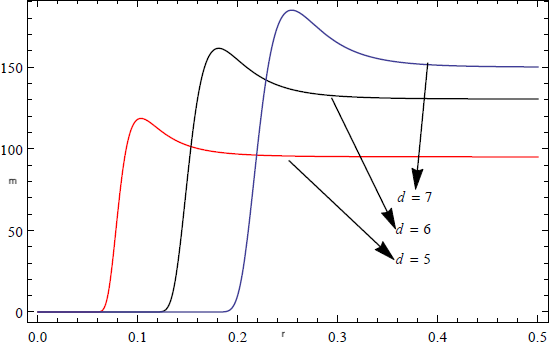}
	\caption{Plot of function $m$ [Eq. (\ref{13})] vs $r_{h}$ for fixed values of $Q=0.01$, $\beta=0.01$ and $l=1$.}\label{Khan1}
\end{figure}
Fig. (\ref{Khan1}) shows that there exists a minimum value $m_{ext}$ of $m$ which corresponds to an extremal black hole. Extremal black holes are the black holes whose horizons coincide, which can be obtained by solving $f(r)=0$ and $df/dr =0$ simultaneously. For $r_h=0$, $m$ takes a finite value $m_0$ for $d\geq3$. If $m_{ext}<m<m_0$, there are two horizons, if $m\geq m_0$, there is a single horizon and if $m<m_{ext}$, there is no event horizon. In Fig. (\ref{Khan2}), the point where curve touches the horizontal axis indicates the position of the event horizon.
 
 In general, the Ricci scalar for the line element (\ref{8}) is 
 \begin{equation}\begin{split}
 R=\frac{(d+1)}{r^2}\bigg(1-f(r)-r\frac{df}{dr}-r^2\frac{d^{2}f}{dr^2}\bigg),\end{split}\label{b}
 \end{equation}
 while the Kretschmann scalar is given by
 \begin{equation}\begin{split}
 K=\frac{(2d+2)\bigg(f(r)-1\bigg)^2}{r^4}+\frac{(d+1)\bigg(\frac{df}{dr}\bigg)^2}{r^2}+\bigg(\frac{d^{2}f}{dr^2}\bigg)^2.\end{split}\label{c}
 \end{equation}
 For arbitrary value of $d$ the metric function (\ref{11}) gives very lengthy and complicated expression for the Ricci scalar which we cannot write here. Avoiding these lengthy calculations we discuss the asymptotic behaviour of curvature invariants for $d=5$.
 Differentiating (\ref{11}) for the 5-dimensional case we can write
 \begin{align} 
 \frac{df}{dr}&=\frac{2r}{l^2}-\frac{2\beta^{\frac{1}{3}}}{r^{\frac{1}{3}}}\bigg[\delta_{d,5}+16m\pi\bigg(\Gamma\bigg(\frac{5}{6},\frac{\beta Q^2}{2r^6}\bigg)-2\Gamma\bigg(\frac{5}{6},\frac{\beta Q^2}{2r^6}\bigg)\bigg)\bigg]^{\frac{1}{3}} \nonumber \\ &-\frac{4}{3^{\frac{2}{3}}r^{\frac{16}{3}}} \bigg[24\pi Q^2\beta^{\frac{5}{6}}\exp{\bigg(\frac{-\beta Q^2}{2r^{6}}\bigg)}\bigg(\Gamma\bigg(\frac{5}{6}\bigg)\bigg)^{-1} -\frac{Q^2\beta}{r^6}\bigg(\Gamma\bigg(\frac{11}{6}\bigg)\bigg)^{-1}\bigg) \nonumber \\ &-(3\delta_{d,5}+48m\pi)\beta^{\frac{5}{6}}r^5-2^{\frac{29}{6}}\pi Q^{\frac{1}{3}}r^5 \bigg(\Gamma\bigg(\frac{5}{6},\frac{\beta Q^2}{2r^6}\bigg)-2\Gamma\bigg(\frac{5}{6},\frac{\beta Q^2}{2r^6}\bigg)\bigg) \bigg] \nonumber \\ 
 &\times \bigg[(3\delta_{d,5}+48m\pi)\beta^{\frac{5}{6}}+2^{\frac{29}{6}}\pi Q^{\frac{1}{3}}\bigg(\Gamma\bigg(\frac{5}{6},\frac{\beta Q^2}{2r^6}\bigg)-2\Gamma\bigg(\frac{5}{6},\frac{\beta Q^2}{2r^6}\bigg)\bigg)\bigg]^{-1}.
 \label{d} 
 \end{align}
 Similarly, the second derivative of the metric function is
 \begin{equation}\begin{split}
 \frac{d^2f}{dr^2}=\frac{2}{l^2}-\frac{2}{3^{\frac{1}{3}}r^{\frac{4}{3}}}\Pi_{1}^{\frac{1}{3}}+\frac{32\pi Q^2\beta^{\frac{2}{9}}\Pi_{2}}{r^{\frac{55}{3}}\Pi_{3}}+\frac{32\Pi_{4}^{2}}{3^\frac{7}{3}r^\frac{4}{3}\Pi_{5}^{\frac{5}{3}}}+\frac{3^{\frac{-4}{3}}r^{\frac{-4}{3}}\Pi_{6}}{\Pi_{5}^{\frac{2}{3}}},
 \label{e}\end{split}\end{equation}
 where 
 \begin{equation}\begin{split}
 \Pi_{1}&=3\delta_{d,5}+48m\pi+2^{\frac{29}{6}}Q^{\frac{1}{3}}\beta^{\frac{-5}{6}}\bigg(\Gamma\bigg(\frac{5}{6},\frac{\beta Q^2}{2r^6}\bigg)-2\Gamma\bigg(\frac{5}{6},\frac{\beta Q^2}{2r^6}\bigg)\bigg),
 \label{f}\end{split}
 \end{equation}
 \begin{equation}
 \Pi_{2}=\exp{\bigg(\frac{-\beta Q^2}{2r^{6}}\bigg)}\bigg[Q^4\beta^2\Gamma\bigg(\frac{5}{6}\bigg)+3r^{12}\Gamma\bigg(\frac{11}{6}\bigg)-Q^2\beta r^6\bigg(5\Gamma\bigg(\frac{5}{6}\bigg)+\Gamma\bigg(\frac{11}{6}\bigg)\bigg)\bigg],\label{g}
 \end{equation}
 \begin{equation}
 \Pi_{3}=\Gamma\bigg(\frac{5}{6}\bigg)\Gamma\bigg(\frac{11}{6}\bigg)\bigg[(3\delta_{d,5}+48m\pi)\beta^{\frac{5}{6}}+2^{\frac{29}{6}}\pi Q^{\frac{1}{3}}\bigg(\Gamma\bigg(\frac{5}{6},\frac{\beta Q^2}{2r^6}\bigg)-2\Gamma\bigg(\frac{5}{6},\frac{\beta Q^2}{2r^6}\bigg)\bigg)\bigg],\label{h}
 \end{equation}
 \begin{align}
 \Pi_{4}&=3\delta_{d,5}+48 m\pi+\frac{84m Q^2\pi}{r}\bigg(\Gamma\bigg(\frac{11}{6}\bigg)\bigg)^{-1}\bigg(\Gamma\bigg(\frac{5}{6}\bigg)\bigg)^{-1}\exp{\bigg(\frac{-\beta Q^2}{2r^{6}}\bigg)} \nonumber  \\ &\times \bigg[Q^2\beta \Gamma\bigg(\frac{5}{6}\bigg)-r^6\Gamma\bigg(\frac{11}{6}\bigg)\bigg]+2^{\frac{29}{6}}\pi Q^{\frac{1}{3}}\beta^{\frac{-5}{6}}\bigg[\Gamma\bigg(\frac{5}{6},\frac{\beta Q^2}{2r^6}\bigg)-2\Gamma\bigg(\frac{5}{6},\frac{\beta Q^2}{2r^6}\bigg)\bigg],\label{i} 
 \end{align}
 \begin{equation}\begin{split}
 \Pi_{5}=3\delta_{d,5}+48m\pi+2^{\frac{29}{6}}\pi Q^{\frac{1}{3}}\beta^{\frac{-5}{6}}\bigg[\Gamma\bigg(\frac{5}{6},\frac{\beta Q^2}{2r^6}\bigg)-2\Gamma\bigg(\frac{5}{6},\frac{\beta Q^2}{2r^6}\bigg)\bigg],\label{j}\end{split}
 \end{equation}
 \begin{align} 
 \Pi_{6}&=96\pi Q^2\exp{\bigg(\frac{-\beta Q^2}{2r^{6}}\bigg)}\bigg[\frac{1}{r^5\Gamma\bigg(\frac{5}{6}\bigg)}-\frac{Q^2\beta}{r^{11}\Gamma\bigg(\frac{5}{6}\bigg)}\bigg]-12\delta_{d,5} \nonumber \\&-192m\pi-2^{\frac{29}{6}}\pi Q^{\frac{1}{3}}\beta^{\frac{-5}{6}}\bigg(\Gamma\bigg(\frac{5}{6},\frac{\beta Q^2}{2r^6}\bigg)-2\Gamma\bigg(\frac{5}{6},\frac{\beta Q^2}{2r^6}\bigg)\bigg).\label{k}  
 \end{align}
 Using the above values in Eq. (\ref{b}), we conclude that $R(r)\rightarrow\infty$ as $r$ tends to zero. This shows that our resulting solution has a true curvature singularity at the origin $r=0$, which indicates that our solution represents a black hole. Furthermore, the metric function $f(r)$ given by (\ref{11}) becomes infinite as $r\rightarrow\infty$, which means that it represents nonasymtotically flat spacetime.

\subsection{Thermodynamics of black holes of DCG with exponential magnetic source} 

The Hawking temperature of the black hole is given by
\begin{align}
T_{H}(r_h)=\frac{r_h}{2\pi l^2}-\frac{r_h}{2\pi}[G(r_h)]^{\frac{1}{d-3}}-\frac{r_h^2}{4\pi}[G(r_h)]^{\frac{2-d}{d-3}}W(r_h),\label{15}
\end{align}
where
\begin{align}
G(r_h)=\frac{16\pi G m}{\Sigma_{d-2}r_h^{d-1}}+\frac{\delta_d}{r_h^{d-1}}+\frac{\pi G Q^{\frac{1}{d-2}}\beta^{\frac{5-2d}{2d-4}}}{(d-2)2^{\frac{21-10d}{2d-4}}r_h^{d-1}}H(r_h),\label{16}
\end{align}
\begin{eqnarray} 
W(r_h)&=&\frac{16\pi G m (1-d)}{r_h^d\Sigma_{d-2}}+\frac{(1-d)\delta_d}{r_h^d}+\frac{\pi G Q^{\frac{1}{d-2}}(1-d)\beta^{\frac{5-2d}{2d-4}}H(r_h)}{(d-2)2^{\frac{21-10d}{2d-4}}r_h^d} \notag \\ &+&\frac{16\pi G Q^2}{(d-2)r_h^{4d-8}}\exp{\bigg(-\frac{\beta Q^2}{2r_h^{2d-4}}\bigg)}\bigg((2d-4)r_h^{2d-4}+(4-2d)\beta^{\frac{3d-7}{2-d}}2^{\frac{9-4d}{2-d}}\bigg),\label{16} 
\end{eqnarray}
and
\begin{align}
H(r_h)=\Gamma{\bigg(\frac{2d-5}{2d-4},\frac{\beta Q^2}{2r_h^{2d-4}}\bigg)}-2\Gamma{\bigg(\frac{4d-9}{2d-4},\frac{\beta Q^2}{2r_h^{2d-4}}\bigg)}.\label{17}
\end{align}

To obtain the expression for heat capacity, we first find the black hole's entropy. A consistent approach which is very fruitful in getting conserved charges of a black hole was developed by Wald \cite{Ar1,Ar2}. Black hole thermodynamics can be developed by using the derived conserved quantities. This Wald's formalism is applicable in general diffeomorphism-invariant theories and even in those types of theories where higher order derivatives are found. Wald's formalism has been helpful in studying thermodynamics of black holes in different theories, for example, Einstein-scalar theory \cite{Ar3,Ar4}, Einstein-Proca theory \cite{Ar5}, Einstein-Yang-Mills theory \cite{Ar6}, gravity with quadratic curvature invariants \cite{Ar7}, Lifschitz gravity \cite{Ar8} etc. In our work too i.e. the case of DCG we use the general formalism of Wald for obtaining the black hole's entropy. Hence the Wald entropy for our solution (\ref{11}) is defined by  
 \begin{eqnarray} 
 S&=&-2\pi\oint d^{n}x\sqrt{h}\frac{\partial L}{\partial R_{\mu\nu\rho\lambda}}\epsilon_{\mu\nu}\epsilon_{\rho\lambda} \notag \\ &=&\frac{(d-3)\Sigma_{d-2}r_h^{d}}{4kG(6-d)} 
\bigg(\frac{k}{r_h^2}+\frac{1}{l^2}\bigg)^{d-3}F_{1}{\bigg[1,\frac{d}{2},\frac{8-d}{2},\frac{-r_h^2}{kl^2}\bigg]},\label{18} 
\end{eqnarray}
where $\epsilon_{\mu\nu}$ is the normal bivector of the $t=const$ and $r=r_h$ hypersurface such that $\epsilon_{\rho\lambda}\epsilon^{\rho\lambda}=-2$, $F_1$ is the Gaussian hypergeometric function. The magnetic potential conjugate to the magnetic charge $Q$ is given by
\begin{equation}
\Phi_m=\frac{\partial m}{\partial Q}=\frac{\Sigma_{d-2}}{(d-2)}\exp{\bigg(-\frac{\beta Q^2}{2r_h^{2d-4}}\bigg)}\bigg[\frac{2 Q}{r_h^{2d-5}}-\frac{2\beta Q^3}{r_h^{4d-9}}\bigg]-\frac{\Sigma_{d-2}Q^{\frac{3-d}{d-2}}\beta^{\frac{5-2d}{2d-4}}H(r_h)}{(d-2)^2 2^{\frac{5-2d}{2d-4}}}.\label{19}
\end{equation}
Using the thermodynamical quantities obtained above, we can now easily verify that the first law of thermodynamics  \cite{28} 
 \begin{equation}
dm=T_H dS+\Phi_m dQ,\label{20}
\end{equation}
is satisfied. Now the specific heat capacity \cite{27,29,30} at a constant charge $Q$ is given by
\begin{equation}
C_{Q}= T_{H}\frac{\partial S}{\partial T_{H}}|_{Q}.\label{21}
\end{equation}
Differentiating (\ref{15}) we get
\begin{eqnarray} 
\frac{\partial T_H}{\partial r_h}&=&\frac{1}{2\pi l^2}-\frac{G(r_h)^{\frac{1}{d-3}}}{2\pi}-\frac{r_h}{2\pi (d-3)G(r_h)^{\frac{d-2}{d-3}}}\frac{dG(r_h)}{dr_h} \notag \\ &-&\bigg(\frac{r_h}{2\pi}G(r_h)^{\frac{d-2}{d-3}}+\frac{(d-2)r_h^2}{4\pi(3-d)G(r_h)^{\frac{2d-5}{d-3}}}
\frac{dG(r_h)}{dr_h}\bigg)W(r_h) \notag \\ 
&-&\frac{r_h^2G(r_h)^{\frac{d-2}{d-3}}}{4\pi}\frac{d W(r_h)}{dr_h},		\label{22}
\end{eqnarray}
where
\begin{eqnarray} 
\frac{d W(r_h)}{d r_h}&=&\frac{16\pi G m d(d-1)}{\Sigma_{d-2}r_h^{d+1}}+\frac{d(d-1)\delta_d}{r_h^{d+1}}+\frac{\pi G d(d-1) Q^{\frac{1}{d-2}}H(r_h)}{(d-2)2^{\frac{21-10d}{2d-4}}\beta^{\frac{2d-5}{2d-4}}r_h^{d+1}} \notag \\ 
&+&\frac{\pi G (1-d)Q^{\frac{1}{d-2}}}{(d-2)2^{\frac{21-10d}{2d-4}}r_h^d\beta^{\frac{2d-5}{2d-4}}}\frac{dH(r_h)}{dr_h}+\frac{16\pi G Q^2(2d-4)^2}{(d-2)r_h^{2d-3}}\exp{\bigg(\frac{-\beta Q^2}{2r_h^{2d-4}}\bigg)} \notag 
\\ &+&\bigg[\frac{8\pi G Q^4\beta(2d-4)^2}{r_h^{6d-11}(d-2)}-\frac{32\pi G Q^2(2d-4)^2}{(d-2)r_h^{4d-7}}\bigg] \notag \\ &\times& \exp{\bigg(\frac{-\beta Q^2}{2r_h^{2d-4}}\bigg)} \bigg(r_h^{2d-4}-\beta^{\frac{3d-7}{2-d}}Q^{\frac{6d-14}{2-d}}2^{\frac{9-4d}{2-d}}\bigg),	\label{23}
\end{eqnarray}
\begin{equation}\begin{split}
\frac{d G(r_h)}{d r_h}=\frac{16\pi G m (1-d)}{\Sigma_{d-2}r_h^d}+\frac{(1-d)\delta_d}{r_h^d}+\frac{\pi G Q^{\frac{1}{d-2}}\beta^{\frac{5-2d}{2d-4}}}{(d-2)r_h^d 2^{\frac{21-10d}{2d-4}}}\bigg((1-d)H(r_h)+r_h\frac{dH(r_h)}{dr_h}\bigg),		
\end{split}\label{24}
\end{equation}
and
\begin{equation}\begin{split}
\frac{d H(r_h)}{d r_h}=\frac{(4-2d)\beta^{\frac{4d-9}{2d-4}}Q^{\frac{4d-9}{d-2}}}{2^{\frac{2d-5}{2d-4}}r_h^{4d-8}}\exp{\bigg(\frac{-\beta Q^2}{2r_h^{2d-4}}\bigg)}\bigg[1-\frac{\beta^{\frac{7-3d}{d-2}}r_h^{2d-4}}{Q^2}\bigg].		
\end{split}\label{25}
\end{equation}
Putting Eqs. (\ref{15}), (\ref{18}) and (\ref{22}) in the general expression of heat capacity (\ref{21}) we obtain
\begin{equation}\begin{split}
C_Q=\bigg(\frac{r_h}{2\pi l^2}-\frac{r_h}{2\pi}[G(r_h)]^{\frac{1}{d-3}}-\frac{r_h^2}{4\pi}[G(r_h)]^{\frac{2-d}{d-3}}W(r_h)\bigg)\frac{(d-3)\Sigma_{d-2}r_h^d Z(r_h)}{4G(6-d)X(r_h)}\bigg(\frac{1}{r_h^2}+
\frac{1}{l^2}\bigg)^{d-3},		
\end{split}\label{26}
\end{equation}
where
\begin{equation}\begin{split}
Z(r_h)=F_1{\bigg[1,\frac{d}{2},\frac{8-d}{2},\frac{-r_h^2}{l^2}\bigg]}\bigg(\frac{d}{r_h}+\frac{(d-3)r_h^2l^2}{r_h^2+l^2}\bigg)-\frac{2r_hd}{l^2(8-d)}
F_1{\bigg[2,\frac{d+2}{2},\frac{10-d}{2},\frac{-r_h^2}{l^2}\bigg]},		
\end{split}\label{27}
\end{equation}
and
\begin{eqnarray} 
X(r_h)&=&\frac{1}{2\pi l^2}-\frac{G(r_h)^{\frac{1}{d-3}}}{2\pi}-\frac{r_h}{2\pi (d-3)G(r_h)^{\frac{d-2}{d-3}}}\frac{dG(r_h)}{dr_h} -\bigg(\frac{r_h}{2\pi}G(r_h)^{\frac{d-2}{d-3}} \notag \\ &+&\frac{(d-2)r_h^2}{4\pi(3-d)G(r_h)^{\frac{2d-5}{d-3}}}
\frac{dG(r_h)}{dr_h}\bigg)W(r_h)-\frac{r_h^2G(r_h)^{\frac{d-2}{d-3}}}{4\pi}\frac{d W(r_h)}{dr_h}.		\label{28}
\end{eqnarray}
The above Eq. (\ref{26}) represents the general expression for black hole's heat capacity for any value of nonlinear electrodynamical parameter $\beta$. The black hole is stable if the Hawking temperature and heat capacity are positive. The black hole becomes unstable in the region where Hawking temperature or heat capacity become negative. The point at which the sign of Hawking temperature changes corresponds to the first order phase transition of black hole. The maximum of Hawking temperature corresponds to the second order phase transition since the heat capacity  at that point is singular.  

\subsection{Hairy black holes of DCG with exponential magnetic source}

The action describing Lovelock-scalar gravity with nonlinear exponential electrodynamics source \cite{32} is defined by
\begin{eqnarray} 
I&=&\frac{1}{16\pi G}\int d^dx\sqrt{-g}\bigg[\sum_{p=0}^{n-1}\frac{\alpha_p}{2^p}\delta^{\mu_1...\mu_{2p}}_{\nu_1...\nu_{2p}}\bigg(a_p R^{\nu_1\nu_2}_{\mu_1\mu_2}...R^{\nu_{2p-1}\nu_{2p}} \notag \\ &+& 16\pi G b_p \phi^{d-4p}S^{\nu_1\nu_2}_{\mu_1\mu_2}...S^{\nu_{2p-1}\nu_{2p}} \bigg)+4\pi G L(\digamma)\bigg],
\label{29} 
\end{eqnarray}
where $L(\digamma)$ corresponds to the Lagrangian density describing the exponential electromagnetic field \cite{21, 32}. The second term in the integrand which contains $\phi^{d-4p}$ denotes the Lagrangian density of the scalar field, where \cite{OR} 
\begin{equation}\begin{split}
S^{\rho\sigma}_{\mu\nu}=\phi^2 R^{\rho\sigma}_{\mu\nu}-2\delta^{[\rho}_{[\mu}\delta^{\sigma]}_{\nu]}\partial_{\xi}\phi\partial^{\xi}\phi-4\phi\delta^{[\rho}_{[\mu}\partial_{\nu]}
\partial^{\sigma]}\phi+48\delta^{[\rho}_{\mu}\partial_{\nu}\phi\partial^{\sigma]}\phi.
\label{30}\end{split}
\end{equation}
Note that, by putting the scalar function $\phi$ equal to zero we get the action function defined in (\ref{1}). The matter tensor corresponding to the scalar field is given by
\begin{equation}\begin{split}
[T^{\nu}_{\mu}]^{(s)}=\sum_{p=0}^{d-3}\frac{b_p}{2^{p+1}}\phi^{d-4p}\delta^{\nu\lambda_1\lambda_2...\lambda_{2p}}_{\mu\rho_1\rho_2...\rho_{2p}}S^{\rho_1\rho_2}
_{\lambda_1\lambda_2}...S^{\rho_{2p-1}\rho_{2p}}_{\lambda_{2p-1}\lambda_{2p}}.
\label{31}\end{split}
\end{equation}
Thus the equation of motion for the scalar field becomes
\begin{equation}\begin{split}
\sum_{p=0}^{d-3}\frac{b_p(d-2p)}{2^{p}}\phi^{d-4p-1}\delta^{\lambda_1\lambda_2...\lambda_{2p}}_{\rho_1\rho_2...\rho_{2p}}S^{\rho_1\rho_2}
_{\lambda_1\lambda_2}...S^{\rho_{2p-1}\rho_{2p}}_{\lambda_{2p-1}\lambda_{2p}}=0.
\label{32}\end{split}
\end{equation}
Taking variation of (\ref{29}) with respect to metric tensor, we get the field equations in the form \cite{B3,B9} 
\begin{equation}
\sum_{p=0}^{n-1}\frac{a_p}{2^{p+1}}\delta^{\nu\lambda_1...\lambda_{2p}}_{\mu\rho_1...\rho_{2p}} R^{\rho_1\rho_2}_{\lambda_1\lambda_2}...R^{\rho_{2p-1}\rho_{2p}}=16\pi G [T^{\nu}_{\mu}]^{(M)}+16\pi G [T^{\nu}_{\mu}]^{(s)},
\label{33}
\end{equation}
where $[T^{\nu}_{\mu}]^{(M)}$ corresponds to the matter tensor of exponential electromagnetic field and $a_{p}$ corresponds to the value defined as in Eq. (\ref{3}) so that the Lovelock gravity becomes DCG. If we take the scalar configuration as \cite{32}
\begin{equation}
\phi=\frac{N}{r},\label{34}
\end{equation}
the equation describing the scalar field becomes
\begin{equation}
\sum_{p=0}^{d-3}\frac{b_p(d-1)!(d^2-d+4p^2)}{(d-2p-1)!}N^{-2p}=0.
\label{35}
\end{equation}
Using the assumption of pure magnetic field and substituting Eqs. (\ref{31}) and (\ref{32}) in Eq. (\ref{33}) we get the metric function in the form
\begin{eqnarray} 
f(r)&=&k+\frac{r^2}{l^2}-r^2\bigg[\frac{16\pi G m}{\Sigma_{d-2}r^{d-1}}+\frac{32\pi Y}{r^d}+\frac{\delta_d}{r^{d-1}} \notag \\ 
&+&\frac{\pi G Q^{\frac{1}{d-2}}\beta^{\frac{5-2d}{2d-4}}}{(d-2)2^{\frac{21-10d}{2d-4}}} \bigg(\Gamma{\bigg(\frac{2d-5}{2d-4},\frac{\beta Q^2}{2r^{2d-4}}\bigg)}-2\Gamma{\bigg(\frac{4d-9}{2d-4},\frac{\beta Q^2}{2r^{2d-4}}\bigg)}\bigg)\bigg]^{\frac{1}{d-3}},\label{36} 
\end{eqnarray}
where we have
\begin{equation}
Y=\sum_{p=0}^{d-3}b_p\frac{(d-2)!}{(d-2p-2)!}N^{d-2p}.\label{37}
\end{equation}
 
 Now we investigate the asymptotic behaviour of metric function (\ref{36}). In what follows, we take $k=1$ and the cases $k=0,-1$ can be studied in similar manners. Thus, for the odd-dimensional spacetimes we obtain 
 \begin{align} 
 f(r)&=1+\frac{r^2}{l^2}-m^{\frac{1}{2s-2}}\bigg[1+\frac{32\pi Y r^{\frac{1}{1-s}}}{2m(s-1)}+\frac{\delta_{d,2s+1}r^{\frac{s-2}{s-1}}}{2m(s-1)}+\frac{\pi G Q^{\frac{1}{2s-1}}\beta^{\frac{3-4s}{4s-2}}r^{\frac{2s^2-s-2}{s-1}}}{(4s-2)(s-1)m 2^{\frac{21-20s-10}{4s-2}}} \nonumber \\ &\times\exp{\bigg(\frac{-\beta Q^2}{2r^{4s-2}}\bigg)}\bigg(\frac{2^{\frac{1}{4s-2}}}{\beta^{\frac{1}{4s-2}}Q^{\frac{1}{2s-1}}}\bigg(2^{4s}-\frac{4s-3}{4s-2}\bigg)r-\frac{2^{\frac{1}{4s-2}}r^{3-4s}}{\beta^{\frac{3-4s}{4s-2}}Q^{\frac{3-4s}{2s-1}}}\bigg)\bigg], \nonumber \\ & \hspace{70mm} d=2s+1, s=1,2,3....
 \label{l} 
 \end{align}
 Similarly, for even-dimensional spacetime we have 
 \begin{align} 
 f(r)&=1+\frac{r^2}{l^2}-m^{\frac{1}{2s-3}}\bigg[1+\frac{32\pi Y r^{\frac{2}{3-2s}}}{2s-3}+\frac{\delta_{d,2s}r^{\frac{2s-5}{2s-3}}}{m(2s-3)}+\frac{\pi G Q^{\frac{1}{2s-3}}\beta^{\frac{5-4s}{4(s-1)}}r^{\frac{4s^2-6s-2}{2s-3}}}{2(s-1)(2s-3)m 2^{\frac{21-10s}{2(s-1)}}} \nonumber \\ &\times\exp{\bigg(\frac{-\beta Q^2}{2r^{4(s-1)}}\bigg)} \bigg(\frac{2^{\frac{1}{4(s-1)}}}{\beta^{\frac{1}{4(s-1)}}Q^{\frac{1}{2(s-1)}}}\bigg(2^{4s-2}-\frac{4s-5}{4s-4}\bigg)r-\frac{2^{\frac{1}{4(s-1)}}r^{5-4s}}{\beta^{\frac{5-4s}{4s-4}}Q^{\frac{5-4s}{2s-2}}}\bigg)\bigg], \nonumber \\ & \hspace{70mm} d=2s, s=1,2,3....
 \label{m} 
 \end{align}
 The above asymptotic expressions of metric functions show that in the case of both odd and even dimensions, metric functions are regular and finite at $r=0$ for all values of $s>1$. This finiteness of metric functions is due to the non-Maxwell behaviour of Lagrangian density (\ref{2}) corresponding to exponential electrodynamics. Furthermore, at $r\rightarrow\infty$ the metric functions become infinite and hence this shows that the spacetime is nonasymptotically flat.
 \begin{figure}[h]
 	\centering
 	\includegraphics[width=0.8\textwidth]{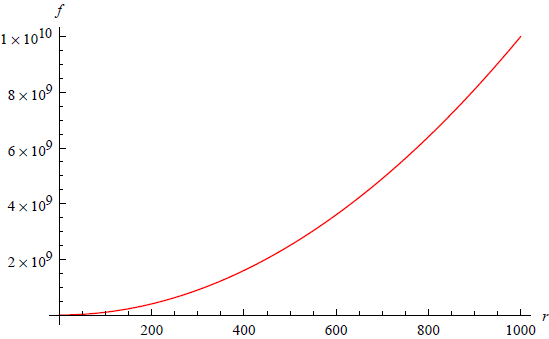}
 	\caption{Plot of function $f(r)$ [Eq. (\ref{36})] vs $r$ for fixed values of $m=10$, $d=6$, $Q=10.0$, $\beta=10.0$ and $l=0.001$.}\label{Khan3}
 \end{figure}
 
 Fig. (\ref{Khan3}) shows the graph of function $f(r)$ vs $r$, for fixed values of $\beta$, $m$, $Q$ and $d$. Note that the values of all other parameters are taken as unity. The point at which the curve touches the horizontal axis indicates the position of event horizon.
 
 Furthermore, by using (\ref{36}) with Eqs. (\ref{b}) and (\ref{c}), one can easily check that both the Ricci scalar and Kretschmann scalar give infinite values at $r=0$ for any value of $d$. This shows that the metric function possesses true curvature singularity at the origin. Hence, the line element (\ref{8}) with metric function given by (\ref{36}) describes a hairy black hole solution of DCG sourced by exponential electrodynamics. For the four-dimensional case, it is seen from Eq. (\ref{35}) that all the coefficients $b_p$ vanish, thus in the case of $d=4$, hairy black holes do not exist. It is easy to see that, by taking $Y=0$, this black hole solution reduces to the black hole with no scalar hair which we derived earlier in Section 2.2

\section{Topological black holes of DCG coupled to power-Yang-Mills theory} 

\subsection{Black hole solution with power-Yang-Mills source}

The action function describing the Lovelock-power-Yang-Mills theory is given by
 \begin{equation}
I=\frac{1}{16\pi G}\int d^dx\sqrt{-g}\bigg[\sum_{p=0}^{n-1}\frac{\alpha_p}{2^p}\delta^{\mu_1...\mu_{2p}}_{\nu_1...\nu_{2p}} R^{\nu_1\nu_2}_{\mu_1\mu_2}...R^{\nu_{2p-1}\nu_{2p}}+(\Upsilon)^q\bigg],\label{38}
\end{equation}
where $\Upsilon$ is the Yang-Mills invariant defined as
\begin{equation}
\Upsilon=\sum_{a=1}^{n(n+1/2)}\bigg(F^{a}_{\lambda\sigma}F^{(a)\lambda\sigma}\bigg), \label{39}
\end{equation}
$q$ is a positive real parameter and the Yang-Mills field is defined by
\begin{equation}
F^{(a)}=dA^{(a)}+\frac{1}{2\eta}C^{(a)}_{(b)(c)}A^{(b)}\wedge A^{(c)},\label{40}
\end{equation}
where, $C^{(a)}_{(b)(c)}$ represents the structure constants of $\frac{n(n+1)}{2}$ -parameter Lie group G, $n=d-2$, $\eta$ denotes the coupling constant and $A^{(a)}$ are the $So(n+1)$ gauge group Yang-Mills potentials. The structure constants have been determined in Ref. \cite{33}. We should keep in mind, that the internal indices $[a,b,...]$ make no difference whether we write them in contravariant or covariant form. Taking variation of action defined in (\ref{38}) with respect to the metric tensor yields the Lovelock field equations (\ref{5}) with matter tensor of Yang-Mills field given by
\begin{equation}
T^{(a)\nu}_{\mu}=-\frac{1}{2}\bigg[\delta^{\nu}_{\mu}\Upsilon^q-4q Tr\bigg(F^{(a)}_{\mu\lambda}F^{(a)\mu\lambda}\bigg)\Upsilon^{q-1}\bigg].\label{41}
\end{equation}
The equations of Yang-Mills field can be obtained if we vary the action with respect to gauge potentials $A^{(a)}$
 \begin{equation} 
d(^{\star}F^{(a)}\Upsilon^{q-1})+\frac{1}{\eta}C^{(a)}_{(b)(c)}\Upsilon^{q-1}A^{(b)}\wedge^{\star}F^{(c)}=0,\label{42}
\end{equation}
where $\star$ denotes the duality of a field. By taking the static metric in the form (\ref{8}), it can easily be checked that power-Yang-Mills equations are satisfied for the choice of Yang-Mills gauge potential one-forms \cite{34,35} defined as
\begin{equation}
A^{(a)}=\frac{Q}{r^2}C^{(a)}_{(i)(j)}x^{i}dx^{j}, r^2=\sum_{i=1}^{d-1}x_{i}^2,\label{43}
\end{equation}
where $Q$ represents the Yang-Mills magnetic charge and $2\leq j+1\leq i\leq d-1$. Thus the symmetric matter tensor (\ref{41}), with
\begin{equation}
\Upsilon=\frac{n(n-1)Q^2}{r^4}=\frac{(d-2)(d-3)Q^2}{r^4},\label{44}
\end{equation}
becomes
\begin{equation}
T^{(a)\nu}_{\mu}=-\frac{1}{2}\Upsilon^q diag[1,1,\zeta,\zeta,...,\zeta], \zeta=\bigg(1-\frac{4q}{(d-2)}\bigg).\label{45}
\end{equation}
The causality principle is satisfied for power-Yang-Mills model if $(\partial L/\partial\Upsilon)\leq0$ and unitarity principle is satisfied when $(\partial L/\partial\Upsilon+2\Upsilon \partial^{2}L/\partial\Upsilon^2) \leq0$ and $\partial^{2}L/\partial\Upsilon^2\geq0$. One can show that the causality does not hold in this model for all values of positive integer $d>3$. However, for $d\leq3$ this principle holds. By using the Lagrangian density of power-Yang-Mills theory it is shown that the unitarity principle holds for all values of $d\geq3$ and $q\leq(1/2)$. When $d\leq3$, this principle is satisfied for all values of $q\geq3$ for power-Yang-Mills magnetic field.

The SEC in DCG coupled to power-Yang-Mills field holds for all values of $q\geq (d-1)/4$. This energy condition tells us that the universe did not experience any acceleration in this model. However WEC and DEC do not hold for all values of $d>3$, in contrast to the exponential model coupled to DCG. Therefore, WEC and DEC hold only for (2+1)-dimensional black holes of DCG with power-Yang-Mills source.
Now by using Eqs. (\ref{8}) and the matter tensor (\ref{45}) in the field equations (\ref{5}), we get
\begin{equation}
\frac{d}{dr}\bigg[r^{d-1}\bigg(\frac{1}{l^2}+\frac{k-f(r)}{r^2}\bigg)^{d-3}\bigg]=\frac{-32\pi G Q^{2q}[(d-2)(d-3)]^q}{r^{4q}}.\label{46}
\end{equation}
Direct integration with respect to $r$ yields the metric function 
\begin{equation}
f(r)=k+\frac{r^2}{l^2}-r^2\bigg[\frac{32\pi G Q^{2q}[(d-2)(d-3)]^q}{(4q-1)r^{4q+d-2}}+\frac{16 \pi G m}{\Sigma_{d-2}r^{d-1}}+\frac{\delta}{r^{d-1}}\bigg]^{\frac{1}{d-3}}.\label{47}
\end{equation}
 Now we discuss the asymptotic behaviour of our solution at $r=0$. For this we will work out for $k=1$ as follows:
 \begin{align} 
 f(r)&=1+\frac{r^2}{l^2}-M r^{\frac{2s-5}{2s-3}}-\frac{32\pi G Q^{2q}(2s-2)^q (2s-3)^{q-1}}{(4q-1)}r^{\frac{4s-8-8qs+12q}{2s-3}} \nonumber \\&+\frac{512\pi^2G^2Q^{4q}(2s-2)^{2q}(2s-3)^{2q-2}}{M(4q-1)^2}r^{\frac{6s-11-16qs+24q}{2s-3}}+O(r^{\frac{8s-15-24qs+36q}{2s-3}}),  \nonumber \\ & \hspace{70mm} d=2s, s=1,2,3...,\label{E} 
 \end{align}
\begin{align} 
f(r)&=1+\frac{r^2}{l^2}-M r^{\frac{s-2}{s-1}}-\frac{32\pi G Q^{2q}(2s-1)^q (2s-2)^{q-1}}{(4q-1)}r^{\frac{4s-6-8qs+8q}{2s-2}} \nonumber \\ &+\frac{512\pi^2G^2Q^{4q}(2s-1)^{2q}(2s-2)^{2q-2}}{M(4q-1)^2}r^{\frac{6s-8-16qs+16q}{2s-2}}+O(r^{\frac{8s-15-24qs+24q}{2s-2}}), \nonumber \\ & \hspace{70mm} d=2s+1, s=1,2,3....\label{F} 
\end{align}
These two expressions indicate that both for even and odd-dimensional spacetimes the metric function (\ref{47}) is regular and finite for all values of $s>1$. Note that in the above we assumed $(m+\delta_{d,2s,2s+1})^{1/d-3}=M$. Furthermore, at $r\rightarrow\infty$ the metric goes to infinity which shows that the spacetime is nonasymptotically flat.

\begin{figure}[h]
	\centering
	\includegraphics[width=0.8\textwidth]{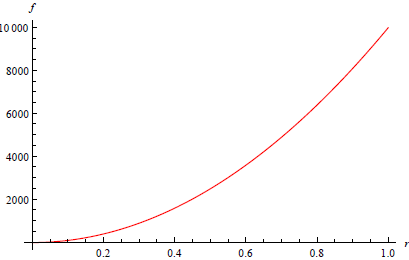}
	\caption{Plot of function $f(r)$ [Eq. (\ref{47})] vs $r$ for fixed values of $m=10$, $d=5$, $Q=0.01$, $q=0.001$ and $l=0.01$.}\label{Khan4}
\end{figure}
Fig. (\ref{Khan4}) shows the graph of $f(r)$ vs $r$ for fixed values of the parameters involved in metric function (\ref{47}). The curve touching the horizontal axis indicates the position of the event horizon. We get extremal black holes when horizons coincide, and they are obtained by solving $f(r)=0$ and $df/dr=0$ simultaneously. This set of simultaneous equations for metric function (\ref{47}), without cosmological constant i.e. for $l\rightarrow\infty$, yields 
\begin{equation}\begin{split}
M_e=\frac{k^{\frac{1}{d-3}}}{r_e^{4d-d^2-1}}-\frac{32\pi G (d-2)^q (d-3)^q Q^{2q}r_e^{1-4q}}{4q-1},\label{AS1}\end{split}
\end{equation}
\begin{equation}\begin{split}
Q_e&=\frac{(d-3)^{1-q}(1-4q)k^{\frac{d-4}{d-3}}r_e^{d^3-9d^2+23d+4q-10}}{16\pi G (d-2)^q (4q+1)} \\ &+\frac{r_e^{d+4q-2}(1-4q)}{32\pi G (d-2)^q(d-3)^q(4q+1)}\bigg(\frac{(1-d)k^{\frac{1}{d-3}}}{r_e^{5d-d^2-2}}\bigg),\label{AS2}\end{split}
\end{equation}
where $M_e=16\pi G m_e/\Sigma_{d-2}+\delta_d$, related to the extremal mass $m_e$, $r_e$ is the extremal radius and $Q_e$ is the extremal charge.
 
Now, we confirm that our resulting solution possesses a central essential singularity. For this, we discuss the asymptotic behaviour of curvature scalars at $r=0$ as follows. Differentiating (\ref{47}) we obtain 
\begin{align} 
\frac{df}{dr}&=\frac{2r}{l^2}-2r\bigg[\frac{16\pi G m}{\Sigma_{d-2}r^{d-1}}+\frac{\delta_d}{r^{d-1}}+\frac{32\pi G Q^{2q}(d-2)^q(d-3)^q}{(4q-1)r^{4q+d-2}}\bigg]^{\frac{1}{d-3}} \nonumber \\ &-\frac{r^2}{d-3}\bigg[\frac{16\pi G m}{\Sigma_{d-2}r^{d-1}}+\frac{\delta_d}{r^{d-1}}+\frac{32\pi G Q^{2q}(d-2)^q(d-3)^q}{(4q-1)r^{4q+d-2}}\bigg]^{\frac{4-d}{d-3}} \nonumber \\ & \times \bigg(\frac{16\pi G m (1-d)}{\Sigma_{d-2}r^{d}}+\frac{(1-d)\delta_d}{r^d}+\frac{32\pi G Q^{2q}(d-2)^q(d-3)^q(2-d-4q)}{(4q-1)r^{4q+d-1}}\bigg).\label{A1} 
\end{align}
Differentiating again we get
\begin{equation}\begin{split}
\frac{d^2f}{dr^2}=\frac{2}{l^2}-2P^{\frac{1}{d-3}}-\frac{4rP^{\frac{4-d}{d-3}}}{d-3}\frac{dP}{dr}-\frac{r^2 P^{\frac{4-d}{d-3}}}{(d-3)}\frac{d^2P}{dr^2}-\frac{r^2 (4-d) P^{\frac{7-2d}{d-3}}}{(d-3)^2}\bigg(\frac{dP}{dr}\bigg)^2,\label{A2}\end{split}
\end{equation}
where
\begin{equation}\begin{split}
P(r)=\frac{16\pi G m}{\Sigma_{d-2}r^{d-1}}+\frac{\delta_d}{r^{d-1}}+\frac{32\pi G Q^{2q}(d-2)^q(d-3)^q}{(4q-1)r^{4q+d-2}},\label{A3}\end{split}
\end{equation}
\begin{equation}\begin{split}
\frac{dP}{dr}=\frac{16\pi G m(1-d)}{\Sigma_{d-2}r^{d}}+\frac{(1-d)\delta_d}{r^{d}}+\frac{32\pi G Q^{2q}(d-2)^q(d-3)^q(2-d-4q)}{(4q-1)r^{4q+d-1}},\label{A4}\end{split}
\end{equation}
and
\begin{equation}\begin{split}
\frac{d^2P}{dr^2}=\frac{16\pi G m d(d-1)}{\Sigma_{d-2}r^{d+1}}+\frac{d(d-1)\delta_d}{r^{d+1}}+\frac{32\pi G Q^{2q}(d-2)^q(d-3)^q(2-d-4q)(1-d-4q)}{(4q-1)r^{4q+d}}.\label{A5}\end{split}
\end{equation}
We can work out the Ricci scalar for any value of $d$ and, therefore, with the help of Eq. (\ref{b}) and the above equations of metric function and its derivatives, we find that $R\rightarrow\infty$ as $r\rightarrow0$. Similarly, Kretschmann scalar (\ref{c}) also possess singularity at $r=0$. Thus the line element (\ref{8}) with the metric function (\ref{47}) represents higher dimensional magnetized black hole solution of DCG for any value of the parameter $q$. However, it indicates that $q=1$ will give the solution of DCG coupled to the standard Yang-Mills theory and for $q=1/4$ the solution does not exist. By substituting the Yang-Mills magnetic charge $Q=0$, we get the neutral solution in DCG. 
     
\subsection{Thermodynamics of black holes of DCG with power-Yang-Mills magnetic source} 

It is observed from the metric function (\ref{47}) that the physical properties of such type of black holes depend on the parameter $q$. The horizons of the black hole are given by $f(r_h)=0$, where $r_h$ represents the location of horizon. Thus
\begin{equation}
m=\frac{\Sigma_{d-2}r_h^{d-1}\bigg(\frac{1}{l^2}+\frac{k}{r_h^2}\bigg)^{d-3}}{16\pi G}-\frac{\Sigma_{d-2}\delta}{16\pi G}-\frac{2Q^{2q}\Sigma_{d-2}[(d-2)(d-3)]^q}{(4q-1)r_h^{4q-1}}.\label{48}
\end{equation}
Equation (\ref{48}) gives the ADM mass of the black hole in terms of horizon radius and Yang-Mills magnetic charge $Q$.
The black hole's Hawking temperature \cite{27} is given by
\begin{eqnarray}
T_{H}&=&\frac{1}{4\pi}\frac{df}{dr}|_{r=r_h} \notag \\ 
&=&\frac{r_h}{2\pi l^2}-\frac{r_h}{2\pi}\bigg[\frac{32\pi G Q^{2q}(d-2)^{q}(d-3)^{q}}{(4q-1)r_h^{4q+d-2}}+\frac{16\pi G m}{\Sigma_{d-2}r_h^{d-1}}+\frac{\delta}{r_h^{d-1}}\bigg]^{\frac{1}{d-3}} \notag 
\\&-&\frac{r_h^2}{4\pi (d-3)}\bigg[\frac{32\pi G Q^{2q}(d-2)^{q}(d-3)^{q}}{(4q-1)r_h^{4q+d-2}}+\frac{16\pi G m}{\Sigma_{d-2}r_h^{d-1}}+\frac{\delta}{r_h^{d-1}}\bigg]^{\frac{4-d}{d-3}} \notag \\ 
&\times& \bigg[\frac{32\pi G(2-d-4q)Q^{2q}(d-2)^{q}(d-3)^{q}}{(4q-1)r_h^{4q+d-1}}+\frac{(1-d)16\pi G m}{\Sigma_{d-2}r_h^{d}}+\frac{(1-d)\delta}{r_h^d}\bigg].\label{49} 
\end{eqnarray}
The entropy of the black hole in this case using Wald's method \cite{Ar1,Ar2} becomes
\begin{equation}\begin{split}
S=\frac{(d-3)\Sigma_{d-2}r_h^{d}}{4kG(6-d}\bigg(\frac{k}{r_h^2}+\frac{1}{l^2}\bigg)^{d-3}F_{1}{\bigg[1,\frac{d}{2},\frac{8-d}{2},\frac{-r_h^2}{kl^2}\bigg]}.
\label{50}\end{split}
\end{equation}
The Yang-Mills magnetic potential conjugate to Yang-Mills magnetic charge $Q$ is given by
\begin{equation}\begin{split}
\Phi_m=\frac{-64\pi G q (d-2)^{q}(d-3)^{q} Q^{2q-1}}{(4q-1)r_h^{4q-1}}.
\label{51}\end{split}
\end{equation}
Thus, using the above thermodynamical quantities, it can be seen that the first law of black hole thermodynamics given by (\ref{20}) is satisfied. The free energy density of the resulting black hole can also be calculated as  \cite{28}
\begin{equation}\begin{split}
\Xi=m-T_H S,
\label{52}\end{split}
\end{equation}
where $m, T_H$ and $S$ are the ADM mass, Hawking temperature and entropy of the black hole given by Eqs. (\ref{48}), (\ref{49}) and (\ref{50}), respectively.
Similarly, the heat capacity at constant charge is given by
\begin{equation}\begin{split}
C_Q=\frac{\bigg[\frac{r_h^{d+1}}{2\pi l^2}-\frac{r_h^{d+1} H_1^{\frac{1}{d-3}}}{2\pi}-\frac{r_h^{d+2}}{4\pi(d-3)H_1^{\frac{d-4}{d-3}}}\frac{dH_1}{dr_h}\bigg](d-3)\Sigma_{d-2}H_2(r_h)\bigg(\frac{k}{r_h^2}
+\frac{1}{l^2}\bigg)^{d-3}}{4k G\bigg[\frac{1}{2\pi l^2}-\frac{H_1^{\frac{1}{d-3}}}{2\pi}-\frac{r_h H_1^{\frac{4-d}{d-3}}}{\pi (d-3)}\frac{dH_1}{dr_h}-\frac{r_h^2H_1^{\frac{4-d}{d-3}}}{4\pi(d-3)}\frac{d^2H_1}{dr_h^2}+\frac{r^2_h(d-4)H_1^{\frac{7-2d}{d-3}}}{4\pi(d-3)^2}\bigg(\frac{dH_1}
{dr_h}\bigg)^2\bigg]},\label{53}\end{split}
\end{equation}
where 
\begin{equation}\begin{split}
H_1(r_h)=\frac{32\pi G Q^{2q}[(d-2)(d-3)]^q}{(4q-1)r_h^{4q+d-2}}+\frac{16\pi G m}{\Sigma_{d-2}r_h^{d-1}}+\frac{\delta}{r_h^{d-1}},\label{54}\end{split}
\end{equation}
\begin{equation} \begin{split}
\frac{dH_1(r_h)}{dr_h}=\frac{32\pi G Q^{2q}(2-4q-d)[(d-2)(d-3)]^q}{(4q-1)r_h^{4q+d-1}}+\frac{16\pi G (1-d)m}{\Sigma_{d-2}r_h^{d}}+\frac{(1-d)\delta}{r_h^{d}},\label{55} \end{split}
\end{equation}
\begin{eqnarray} 
\frac{d^2H_1(r_h)}{dr_h^2}&=&\frac{32\pi G Q^{2q}(2-4q-d)(1-4q-d)[(d-2)(d-3)]^q}{(4q-1)r_h^{4q+d}} \notag \\ 
&+&\frac{16\pi G d(d-1)m}{\Sigma_{d-2}r_h^{d+1}}+\frac{d(d-1)\delta}{r_h^{d+1}},\label{55} 
\end{eqnarray}
and
\begin{equation}\begin{split}
H_2(r_h)=\bigg[\frac{d}{r_h}+\frac{r_h^2 l^2 (d-3)}{kl^2+r_h^2}\bigg]F_1{\bigg(1,\frac{d}{2},\frac{8-d}{2},\frac{-r_h^2}{kl^2}\bigg)}+\frac{2r_hd}{kl^2(d-2)}F_1{\bigg(2,\frac{d+2}{2},\frac{10-d}{2},
\frac{-r_h^2}{kl^2}\bigg)}
.\label{54}\end{split}
\end{equation}

\subsection{Hairy black holes of DCG with power-Yang-Mills magnetic source} 

The action function for Lovelock-scalar gravity with power-Yang-Mills source \cite{32} is defined by
\begin{align} 
I&=\frac{1}{16\pi G}\int d^dx\sqrt{-g}\bigg[\sum_{p=0}^{n-1}\frac{\alpha_p}{2^p}\delta^{\mu_1...\mu_{2p}}_{\nu_1...\nu_{2p}}\bigg(a_p R^{\nu_1\nu_2}_{\mu_1\mu_2}...R^{\nu_{2p-1}\nu_{2p}} \nonumber \\ &+16\pi G b_p \phi^{d-4p}S^{\nu_1\nu_2}_{\mu_1\mu_2}...S^{\nu_{2p-1}\nu_{2p}}\bigg)+4\pi G (\Upsilon)^q\bigg],
\label{55} 
\end{align}
where $(\Upsilon)$ corresponds to the power-Yang-Mills invariant defined by (\ref{39}) and (\ref{40}) and we have used the Lagrangian density of the scalar field given by
\begin{equation}
L_s=\sum_{p=0}^{n-1}\frac{b_p}{2^p}\phi^{d-4p}\delta^{\mu_1...\mu_{2p}}_{\nu_1...\nu_{2p}}S^{\nu_1\nu_2}_{\mu_1\mu_2}...S^{\nu_{2p-1}\nu_{2p}},\label{56}
\end{equation}
where $S^{\rho\sigma}_{\mu\nu}$ is given by Eq. (\ref{30}). Note that, by putting scalar function $\phi$ equal to zero, the above action function reduces to the case of DCG coupled to power-Yang-Mills field, i.e., (\ref{38}).

Taking variation of (\ref{55}) with respect to the metric tensor, we get the field equations as 
\begin{equation}
\sum_{p=0}^{n-1}\frac{a_p}{2^{p+1}}\delta^{\nu\lambda_1...\lambda_{2p}}_{\mu\rho_1...\rho_{2p}} R^{\rho_1\rho_2}_{\lambda_1\lambda_2}...R^{\rho_{2p-1}\rho_{2p}}=16\pi G [T^{\nu}_{\mu}]^{(M)}+16\pi G [T^{\nu}_{\mu}]^{(s)},
\label{57}
\end{equation}
where $[T^{\nu}_{\mu}]^{(M)}$ corresponds to the matter tensor of power-Yang-Mills field while $a_{p}$ corresponds to the value defined as in Eq. (\ref{3}) so that the Lovelock gravity becomes DCG. The stress-energy tensor corresponding to the scalar field is given by (\ref{31}) and the equation of motion of scalar field is given by (\ref{32}). Choosing the scalar configuration (\ref{34}) and using Eqs. (\ref{31}) and (\ref{32}) in Eq. (\ref{57}), we get the metric function in the form 

\begin{equation}\begin{split}
f(r)=k+\frac{r^2}{l^2}-r^2\bigg[\frac{16\pi G m}{\Sigma_{d-2}r^{d-1}}+\frac{32\pi Y}{r^d}+\frac{\delta_d}{r^{d-1}}+\frac{32\pi G Q^{2q}(d-2)^q(d-3)^q}{(4q-1)r^{4q+d-2}}\bigg]^{\frac{1}{d-3}},\label{58}\end{split}
\end{equation}
where $Y$ is given by (\ref{37}).

Now, we discuss the asymptotic behaviour of metric function (\ref{58}) at $r=0$. So, assuming $M=16\pi Gm/\Sigma_{d-2}$ and $\mu=\delta_d+M$ we can write the asymptotic expressions as follows:

\begin{align} 
f(r)&=1+\frac{r^2}{l^2}-(32\pi Y)^{\frac{1}{2s-3}}\bigg[r^{\frac{2s-6}{2s-3}}+\frac{\mu r^{\frac{4s-9}{2s-3}}}{32\pi Y (2s-3)} \nonumber \\ &+\frac{G Q^{2q}(2s-2)^q (2s-3)^{q-1}}{Y(4q-1)}r^{\frac{6s-12-8qs+12q}{2s-3}}+O(r^{\frac{6s-12}{2s-3}})\bigg], d=2s, s=1,2,3...,\label{I} 
\end{align}

\begin{align} 
f(r)&=1+\frac{r^2}{l^2}-(32\pi Y)^{\frac{1}{2s-2}}\bigg[r^{\frac{2s-5}{2s-2}}+\frac{\mu r^{\frac{4s-7}{2s-3}}}{32\pi Y (2s-2)} \nonumber \\ &+\frac{G Q^{2q}(2s-1)^q (2s-2)^{q-1}}{Y(4q-1)}r^{\frac{6s-9-8qs+8q}{2s-2}}+O(r^{\frac{6s-9}{2s-2}})\bigg], d=2s+1, s=1,2,3....\label{J} 
\end{align} 

The above asymptotic expansions show that for the even-dimensional spacetime, the metric function is finite and regular for all values of $s$ at the origin. However, for the case of odd dimensions, the metric function is regular and finite at the origin when $s>1$. Note that this finiteness is due to the nonlinear behaviour of Lagrangian density corresponding to power-Yang-Mills theory. Furthermore, the solution given by (\ref{58}) is nonasymptotically flat as it grows infinitely for large values of $r$. 

\begin{figure}[h]
	\centering
	\includegraphics[width=0.8\textwidth]{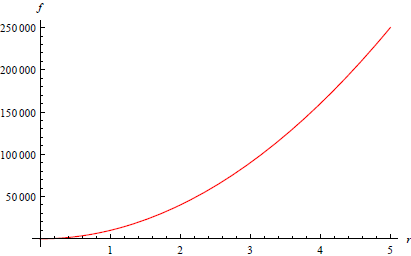}
	\caption{Plot of function $f(r)$ [Eq. (\ref{58})] vs $r$ for fixed values of $m=100$, $d=5$, $Q=0.10$, $q=0.001$ and $l=0.01$.}\label{Khan5}
\end{figure} 

Fig. (\ref{Khan5}) shows the graph of $f(r)$ vs $r$, for fixed values of parameters involved in metric function (\ref{58}). The event horizon is the value of $r$ at which the curve meets the horizontal axis. Extremal quantities corresponding to extremal black hole solution \cite{25} can be obtained by solving $f(r)=0$ and $df/dr=0$ simultaneously. This set of simultaneous equations for our metric function (\ref{58}), without cosmological constant i.e. for $l\rightarrow\infty$, is given by 
\begin{equation}\begin{split}
M_e=\frac{k^{\frac{1}{d-3}}}{r_e^{4d-d^2-1}}-\frac{32\pi Y}{r_e}-\frac{32\pi G (d-2)^q (d-3)^q Q^{2q}r_e^{1-4q}}{4q-1},\label{AS3}\end{split}
\end{equation}
\begin{align}
Q_e&=\frac{(d-3)^{1-q}(1-4q)k^{\frac{d-4}{d-3}}r_e^{d^3-9d^2+23d+4q-10}}{16\pi G (d-2)^q (4q+1)} \nonumber \\ &+\frac{r_e^{d+4q-2}(1-4q)}{32\pi G (d-2)^q(d-3)^q(4q+1)}\bigg(\frac{(1-d)k^{\frac{1}{d-3}}}{r_e^{5d-d^2-2}}-\frac{32\pi Y}{r_e^d}\bigg),\label{AS4} 
\end{align}
where $M_e= 16\pi G m_e /\Sigma_{d-2}+\delta_d$, related to the extremal mass $m_e$, $r_e$ is the extremal radius and $Q_e$ is the extremal charge.

Now we confirm that our resulting solution represents a black hole. This can be done by studying the asymptotic behaviour of curvature invariants at the origin. For this we differentiate Eq. (\ref{58}) and obtain 
\begin{align} 
\frac{df}{dr}&=\frac{2r}{l^2}-2r\bigg[\frac{16\pi G m}{\Sigma_{d-2}r^{d-1}}+\frac{32\pi Y}{r^d}+\frac{\delta_d}{r^{d-1}}+\frac{32\pi G Q^{2q}(d-2)^q(d-3)^q}{(4q-1)r^{4q+d-2}}\bigg]^{\frac{1}{d-3}} \nonumber \\ &-\frac{r^2}{d-3}\bigg[\frac{16\pi G m}{\Sigma_{d-2}r^{d-1}}+\frac{32\pi Y}{r^d}+\frac{\delta_d}{r^{d-1}}+\frac{32\pi G Q^{2q}(d-2)^q(d-3)^q}{(4q-1)r^{4q+d-2}}\bigg]^{\frac{4-d}{d-3}} \nonumber \\ &\times \bigg(\frac{16\pi G m (1-d)}{\Sigma_{d-2}r^{d}}-\frac{32\pi d Y}{r^{d+1}}+\frac{(1-d)\delta_d}{r^d}+\frac{32\pi G Q^{2q}(d-2)^q(d-3)^q(2-d-4q)}{(4q-1)r^{4q+d-1}}\bigg). \label{A6} 
\end{align}
Similarly, by differentiating again we get
\begin{equation}\begin{split}
\frac{d^2f}{dr^2}=\frac{2}{l^2}-2P^{\frac{1}{d-3}}-\frac{4rP^{\frac{4-d}{d-3}}}{d-3}\frac{dP}{dr}-\frac{r^2 P^{\frac{4-d}{d-3}}}{(d-3)}\frac{d^2P}{dr^2}-\frac{r^2 (4-d) P^{\frac{7-2d}{d-3}}}{(d-3)^2}\bigg(\frac{dP}{dr}\bigg)^2,\label{A7}\end{split}
\end{equation}
where
\begin{equation}\begin{split}
P(r)=\frac{16\pi G m}{\Sigma_{d-2}r^{d-1}}+\frac{32\pi Y}{r^d}+\frac{\delta_d}{r^{d-1}}+\frac{32\pi G Q^{2q}(d-2)^q(d-3)^q}{(4q-1)r^{4q+d-2}},\label{A8}\end{split}
\end{equation}
\begin{equation}\begin{split}
\frac{dP}{dr}=\frac{16\pi G m(1-d)}{\Sigma_{d-2}r^{d}}-\frac{32\pi d Y}{r^{d+1}}+\frac{(1-d)\delta_d}{r^{d}}+\frac{32\pi G Q^{2q}(d-2)^q(d-3)^q(2-d-4q)}{(4q-1)r^{4q+d-1}},\label{A9}\end{split}
\end{equation}
and
\begin{align} 
\frac{d^2P}{dr^2}&=\frac{16\pi G m d(d-1)}{\Sigma_{d-2}r^{d+1}}+\frac{32\pi d(d+1)Y}{r^{d+2}}+\frac{d(d-1)\delta_d}{r^{d+1}} \nonumber \\ &+\frac{32\pi G Q^{2q}(d-2)^q(d-3)^q(2-d-4q)(1-d-4q)}{(4q-1)r^{4q+d}}.\label{A10} 
\end{align}
 Thus using Eqs. (\ref{58}), (\ref{A6}) and (\ref{A7}) and the general expression of Ricci scalar ({\ref{b}}), one can easily check that in the limit $r\rightarrow0$ Ricci scalar becomes infinite. Thus, we conclude that metric function (\ref{58}) represents an object which has true curvature singularity at $r=0$. Therefore, line element (\ref{8}) with metric function given by (\ref{58}) represents a hairy black hole solution of DCG in the background of power-Yang-Mills field. Hence, we derive a large family of black hole solutions for any value of real parameter $q$ except for the case $q=1/4$. It is easy to see that, by taking $Y=0$, this black hole solution reduces to the black hole with no scalar hair which we derived in Section 3.1.

\section{Black holes of DCG and Yang-Mills hierarchies} 

In this section we study the possible black holes of DCG whose gravitational field is sourced by the superposition of different power-Yang-Mills field. The Yang-Mills hierarchies in diverse dimensions have been discussed in the literature \cite{36}. Here we begin with an action defined as
 \begin{equation}
I=\frac{1}{16\pi G}\int d^dx\sqrt{-g}\bigg[\sum_{p=0}^{n-1}\frac{\alpha_p}{2^p}\delta^{\mu_1...\mu_{2p}}_{\nu_1...\nu_{2p}} R^{\nu_1\nu_2}_{\mu_1\mu_2}...R^{\nu_{2p-1}\nu_{2p}}+\sum_{j=0}^{q}c_j(\Upsilon)^j\bigg],
\label{59}
\end{equation}
where $\Upsilon$ represents the Yang-Mills invariant defined in (\ref{39}), $c_j, j\geq1$ is a coupling constant. The variation of the above action with respect to the metric tensor gives (\ref{5}) with the matter tensor given by 
\begin{equation}
T^{\nu}_{\mu}=-\frac{1}{2}\sum_{j=0}^{q}c_j\bigg(\delta^{\nu}_{\mu}\Upsilon^j-4j Tr(F^{(a)\nu\sigma}_{\mu\sigma})\bigg).
\label{60}
\end{equation}
The Yang-Mills equations are determined by varying (\ref{59}) with respect to the gauge potential $A^{(a)}$
\begin{equation}
\sum_{j=0}^{q}c_j\bigg[d(^{\star}F^{(a)}\Upsilon^{j-1})+\frac{1}{\eta}C^{(a)}_{(b)(c)}\Upsilon^{j-1}A^{(b)}\wedge ^{\star}F^{(c)}\bigg]=0.
\label{61}
\end{equation}
Our $d=n+2$-dimensional line element ansatz is given by (\ref{8}) and the power-Yang-Mills field ansatz would be chosen as before such that the matter tensor takes the form
\begin{equation}
T^{\nu}_{\mu}=-\frac{1}{2}\sum_{j=0}^{q}c_j \Upsilon^j diag[1,1,\xi,\xi,...,\xi],
\label{62}
\end{equation}
where $\xi=1-\frac{4j}{(d-2)}$. Thus using Eqs. (\ref{3}), (\ref{8}) and (\ref{62}) in (\ref{5}) we get the metric function
\begin{equation}\begin{split}
f(r)=k+\frac{r^2}{l^2}-r^2\bigg[32\pi G \sum_{j=0}^{q}\frac{(d-2)^j(d-3)^jQ^{2j}}{(4j-1)r^{4j+d-2}}+\frac{16\pi m G}{\Sigma_{d-2}r^{d-1}}+\frac{\delta}{r^{d-1}}\bigg]^{\frac{1}{d-3}}.\label{63}\end{split}
\end{equation}
The above equation indicates that for $j=0$ the neutral black hole solution is obtained, for $j=4$ the solution is undefined and for taking a unique value of $j=q$ we get the solution obtained in Section 3.1. The ADM mass in terms of the outer horizon radius $r_h$ is 
\begin{equation}
m=\frac{\Sigma_{d-2}r_h^{d-1}\bigg(\frac{1}{l^2}+\frac{k}{r_h^2}\bigg)^{d-3}}{16\pi G}-\frac{\Sigma_{d-2}\delta}{16\pi G}-\sum_{j=0}^{q}\frac{2Q^{2j}\Sigma_{d-2}[(d-2)(d-3)]^j}{(4j-1)r_h^{4j-1}}.\label{64}
\end{equation}
When superposition of different power-Yang-Mills sources are taken into account, the metric function corresponding to the hairy black hole solution of DCG takes the form
\begin{equation}\begin{split}
f(r)=k+\frac{r^2}{l^2}-r^2\bigg[32\pi G \sum_{j=0}^{q}\frac{(d-2)^j(d-3)^jQ^{2j}}{(4j-1)r^{4j+d-2}}+\frac{16\pi m G}{\Sigma_{d-2}r^{d-1}}+\frac{\delta}{r^{d-1}}+\frac{32\pi Y}{r^d}\bigg]^{\frac{1}{d-3}},\label{65}\end{split}
\end{equation}
where $Y$ is given by (\ref{37}). The asymptotic behaviour of these solutions is similar to those obtained earlier in the case of DCG coupled to power-Yang-Mills source. Also, one can easily check that for metric functions (\ref{63}) and (\ref{65}), the curvature invariants possess an essential singularity at the origin $r=0$ which is the mathematical aspect of the black hole.

\section{Summary and conclusion} 

In this paper, new magnetized black hole solutions of DCG are constructed. In order to do this we studied black holes of DCG in the framework of both exponential electrodynamics and power-Yang-Mills theory. First, we derived both topological black hole solutions and hairy black hole solutions of DCG with pure exponential magnetic source. These solutions depend on the parameter $\beta$ of exponential electrodynamics. For the model of exponential electrodynamics, there is no need to impose the condition on the matter tensor for making it traceless, so we conclude that the scale and dual invariances are completely violated here. Further, the components of the matter tensor obtained from the Lagrangian density of this model satisfy all the energy conditions along with causality and unitarity principles in some specified region of the radial coordinate. For any value of parameter $\beta$, it is possible to obtain a solution which is regular at the origin. Moreover, these solutions are nonasymptotically flat for nonzero value of constant $l$ while in the limit $l$ approaches to infinity one can get the asymptotically flat solutions. It is shown that the metric functions are finite at the origin; this finiteness property of metric functions is due to the nonlinear behaviour of electromagnetic field characterized by Lagrangian density (\ref{2}). Second, we use a model of power-Yang-Mills theory and derive a large family of topological black hole solutions in DCG. These solutions depend on the parameter $q$ and are also nonasymptotically flat for any nonzero value of $l$. The case $q=1$ gives the solutions of black holes with standard Yang-Mills source and for $q=1/4$ the solution does not exist. The hairy black hole solutions are also derived in the framework of power-Yang-Mills theory which are reducible to black holes with no scalar hair for $Y=0$ and to neutral black holes for $Q=0$.

The thermodynamics of topological black holes is also studied within both the exponential electrodynamics and power-Yang-Mills theory. Thermodynamical quantities such as entropy, Hawking temperature and specific heat capacity at constant charge of resulting magnetized higher dimensional black holes of DCG have been worked out. The first law of black hole thermodynamics has also been shown to hold for these black hole solutions. 

\section*{Acknowledgements}
Research grants from the Higher Education Commission of Pakistan under its Projects No. 20-2087 and No. 6151 are gratefully acknowledged.

\end{document}